\documentclass[aps,prd,twocolumn,showpacs,preprintnumbers,superscriptaddress,tightenlines,floatfix]{revtex4}

\usepackage{amsmath}
\usepackage{amssymb}
\usepackage{graphicx}
\usepackage{epsfig}
\usepackage{color}
\usepackage{multirow}

\newcommand{\re}{{{\cal R}e}}
\newcommand{\im}{{{\cal I}m}}

\newcommand{\rez}{{{\cal R}e(z)}}
\newcommand{\imz}{{{\cal I}m(z)}}
\newcommand{\dgog}{{\Delta\Gamma_d/\Gamma_d}}
\newcommand{\taubz}{{\tau_\bz}}
\newcommand{\taubp}{{\tau_\bp}}
\newcommand{\dmd}{{\Delta m_d}}

\newcommand{\mev}{{\rm MeV}}

\newcommand{\mevcc}{{\rm MeV}/c^2}
\newcommand{\gev}{{\rm GeV}}
\newcommand{\gevc}{{\rm GeV}/c}
\newcommand{\gevcc}{{\rm GeV}/c^2}

\newcommand{\frec}{{f_{\rm rec}}}
\newcommand{\ftag}{{f_{\rm tag}}}
\newcommand{\fcp}{{f_{CP}}}
\newcommand{\trec}{{t_{\rm rec}}}
\newcommand{\ttag}{{t_{\rm tag}}}

\newcommand{\Dz}{{\Delta Z}}
\newcommand{\Dt}{{\Delta t}}
\newcommand{\qtag}{{q_{\rm tag}}}

\newcommand{\ups}{{\Upsilon(4S)}}
\newcommand{\bz}{{B^0}}
\newcommand{\bb}{{\overline{B}{}^0}}

\newcommand{\db}{{\overline{D}{}^0}}
\newcommand{\bp}{{B^+}}

\newcommand{\Dm}{{D^-}}
\newcommand{\jpsi}{{J/\psi}}
\newcommand{\ks}{{K_S^0}}
\newcommand{\kl}{{K_L^0}}
\newcommand{\kz}{{K^0}}
\newcommand{\pip}{{\pi^+}}
\newcommand{\pim}{{\pi^-}}
\newcommand{\elp}{{e^+}}
\newcommand{\elm}{{e^-}}
\newcommand{\kp}{{K^+}}

\newcommand{\lcp}{{\lambda_{CP}}}
\newcommand{\etacplcp}{{\eta_{CP}\lambda_{CP}}}

\begin{document}

\title{\boldmath Search for Time-Dependent $CPT$ Violation\\in Hadronic and Semileptonic $B$ Decays}

\affiliation{University of Bonn, Bonn}
\affiliation{Budker Institute of Nuclear Physics SB RAS and Novosibirsk State University, Novosibirsk 630090}
\affiliation{Faculty of Mathematics and Physics, Charles University, Prague}
\affiliation{University of Cincinnati, Cincinnati, Ohio 45221}
\affiliation{Department of Physics, Fu Jen Catholic University, Taipei}
\affiliation{Justus-Liebig-Universit\"at Gie\ss{}en, Gie\ss{}en}
\affiliation{Gifu University, Gifu}
\affiliation{Gyeongsang National University, Chinju}
\affiliation{Hanyang University, Seoul}
\affiliation{University of Hawaii, Honolulu, Hawaii 96822}
\affiliation{High Energy Accelerator Research Organization (KEK), Tsukuba}
\affiliation{Hiroshima Institute of Technology, Hiroshima}
\affiliation{Indian Institute of Technology Guwahati, Guwahati}
\affiliation{Indian Institute of Technology Madras, Madras}
\affiliation{Institute of High Energy Physics, Chinese Academy of Sciences, Beijing}
\affiliation{Institute of High Energy Physics, Vienna}
\affiliation{Institute of High Energy Physics, Protvino}
\affiliation{Institute of Mathematical Sciences, Chennai}
\affiliation{Institute for Theoretical and Experimental Physics, Moscow}
\affiliation{J. Stefan Institute, Ljubljana}
\affiliation{Kanagawa University, Yokohama}
\affiliation{Institut f\"ur Experimentelle Kernphysik, Karlsruher Institut f\"ur Technologie, Karlsruhe}
\affiliation{Korea Institute of Science and Technology Information, Daejeon}
\affiliation{Korea University, Seoul}
\affiliation{Kyungpook National University, Taegu}
\affiliation{\'Ecole Polytechnique F\'ed\'erale de Lausanne (EPFL), Lausanne}
\affiliation{Faculty of Mathematics and Physics, University of Ljubljana, Ljubljana}
\affiliation{University of Maribor, Maribor}
\affiliation{Max-Planck-Institut f\"ur Physik, M\"unchen}
\affiliation{University of Melbourne, School of Physics, Victoria 3010}
\affiliation{Graduate School of Science, Nagoya University, Nagoya}
\affiliation{Kobayashi-Maskawa Institute, Nagoya University, Nagoya}
\affiliation{Nara Women's University, Nara}
\affiliation{National Central University, Chung-li}
\affiliation{National United University, Miao Li}
\affiliation{Department of Physics, National Taiwan University, Taipei}
\affiliation{H. Niewodniczanski Institute of Nuclear Physics, Krakow}
\affiliation{Nippon Dental University, Niigata}
\affiliation{Niigata University, Niigata}
\affiliation{University of Nova Gorica, Nova Gorica}
\affiliation{Osaka City University, Osaka}
\affiliation{Pacific Northwest National Laboratory, Richland, Washington 99352}
\affiliation{Research Center for Nuclear Physics, Osaka University, Osaka}
\affiliation{RIKEN BNL Research Center, Upton, New York 11973}
\affiliation{University of Science and Technology of China, Hefei}
\affiliation{Seoul National University, Seoul}
\affiliation{Sungkyunkwan University, Suwon}
\affiliation{School of Physics, University of Sydney, NSW 2006}
\affiliation{Tata Institute of Fundamental Research, Mumbai}
\affiliation{Excellence Cluster Universe, Technische Universit\"at M\"unchen, Garching}
\affiliation{Toho University, Funabashi}
\affiliation{Tohoku Gakuin University, Tagajo}
\affiliation{Tohoku University, Sendai}
\affiliation{Department of Physics, University of Tokyo, Tokyo}
\affiliation{Tokyo Institute of Technology, Tokyo}
\affiliation{Tokyo Metropolitan University, Tokyo}
\affiliation{Tokyo University of Agriculture and Technology, Tokyo}
\affiliation{CNP, Virginia Polytechnic Institute and State University, Blacksburg, Virginia 24061}
\affiliation{Yamagata University, Yamagata}
\affiliation{Yonsei University, Seoul}
  \author{T.~Higuchi}\affiliation{High Energy Accelerator Research Organization (KEK), Tsukuba} 
  \author{K.~Sumisawa}\affiliation{High Energy Accelerator Research Organization (KEK), Tsukuba} 
  \author{I.~Adachi}\affiliation{High Energy Accelerator Research Organization (KEK), Tsukuba} 
  \author{H.~Aihara}\affiliation{Department of Physics, University of Tokyo, Tokyo} 
  \author{D.~M.~Asner}\affiliation{Pacific Northwest National Laboratory, Richland, Washington 99352} 
  \author{V.~Aulchenko}\affiliation{Budker Institute of Nuclear Physics SB RAS and Novosibirsk State University, Novosibirsk 630090} 
  \author{T.~Aushev}\affiliation{Institute for Theoretical and Experimental Physics, Moscow} 
  \author{A.~M.~Bakich}\affiliation{School of Physics, University of Sydney, NSW 2006} 
  \author{A.~Bay}\affiliation{\'Ecole Polytechnique F\'ed\'erale de Lausanne (EPFL), Lausanne} 
  \author{K.~Belous}\affiliation{Institute of High Energy Physics, Protvino} 
% ...
\collaboration{The Belle Collaboration}

\begin{abstract}
We report a new sensitive search for $CPT$ violation, which includes
improved measurements of the $CPT$-violating parameter $z$ and
the total decay-width difference normalized to the averaged width $\dgog$ of the two $B_d$ mass eigenstates.
The results are based on
a data sample of $535\times 10^6$ $B\overline{B}$ pairs collected at the $\Upsilon(4S)$ resonance
with the Belle detector at the KEKB asymmetric-energy $e^+e^-$ collider.
We obtain
$\rez  = [+1.9\pm3.7 (\rm {stat}) \pm3.3 (\rm {syst})]\times10^{-2}$,
$\imz  = [-5.7\pm3.3 (\rm {stat}) \pm3.3 (\rm {syst})]\times10^{-3}$, and
$\dgog = [-1.7\pm1.8 (\rm {stat}) \pm1.1 (\rm {syst})]\times10^{-2}$,
all of which are consistent with zero.
This is the most precise single measurement of these parameters
in the neutral $B$-meson system to date.
\end{abstract}


\maketitle

\par

$CPT$ invariance is one of the most fundamental theoretical concepts;
its violation would have a serious impact on physics in general,
and would require new physics beyond the standard model (SM).
$CPT$ violation requires the breakdown of some fundamental
underlying physical assumption in the new physics beyond the SM,
for example, violation of Lorentz invariance~\cite{bib:PRL.89.231602}.
Several searches for $CPT$ violation have been carried out;
for example, the Belle and {\it BaBar} collaborations have published
measurements of $CPT$-violating parameters in the neutral $B$-meson
system~[\onlinecite{bib:Belle-CPT-Dilepton}-\onlinecite{bib:BaBar-CPT-Dilepton}],
and the CPLEAR, KLOE, and KTeV collaborations have done
so in the neutral $K$-meson system~[\onlinecite{bib:CPLEAR-CPT}-\onlinecite{bib:KTeV-CPT}].

\par

In the presence of $CPT$ violation,
the flavor and mass eigenstates of the neutral $B$ mesons are related by
$|B_L\rangle = p\sqrt{1-z}|\bz\rangle+q\sqrt{1+z}|\bb\rangle$ and
$|B_H\rangle = p\sqrt{1+z}|\bz\rangle-q\sqrt{1-z}|\bb\rangle$,
where $|B_L\rangle~(|B_H\rangle)$ is a light (heavy) mass eigenstate.
Here $z$ is a complex parameter accounting for $CPT$ violation; $CPT$ is violated if $z\ne0$.
In the decay chain $\ups\to\bz\bb\to\frec\ftag$,
where one of the $B$-mesons decays at time $\trec$ to a reconstructed final state $\frec$
and the other decays at time $\ttag$ to a final state $\ftag$ that distinguishes between $\bz$ and $\bb$,
the general time-dependent decay rate with $CPT$ violation allowed is given by~\cite{bib:BaBar-CPT-Hadronic}
\begin{widetext}
\begin{eqnarray}
\lefteqn{{\cal P}(\Dt; \frec\ftag)
	\;=\; \frac{\Gamma_d}{2}e^{-\Gamma_d|\Delta t|} \biggl[ \,\, \biggr.
	  \frac{|\eta_+|^2+|\eta_-|^2}{2}\cosh \left(\frac{\Delta\Gamma_d}{2}\Delta t\right)
		-          \re(\eta_+^*\eta_-)   \sinh \left(\frac{\Delta\Gamma_d}{2}\Delta t\right)} \nonumber \\
		&&\qquad\qquad\qquad\qquad\qquad\qquad
		+ \frac{|\eta_+|^2-|\eta_-|^2}{2}\cos  \left(\dmd\Dt\right)
		+         \im(\eta_+^*\eta_-)    \sin  \left(\dmd\Dt\right)
		\biggl.\biggr], \label{eq:master_PDF} \\
\eta_+ &\equiv& {\cal A}_{\bz\to\frec} {\cal A}_{\bb\to\ftag} - {\cal A}_{\bb\to\frec} {\cal A}_{\bz\to\ftag}, \label{eq:etap} \\
\eta_- &\equiv& \sqrt{1-z^2}\left(\frac{p}{q} {\cal A}_{\bz\to\frec} {\cal A}_{\bz\to\ftag} - \frac{q}{p} {\cal A}_{\bb\to\frec} {\cal A}_{\bb\to\ftag}\right) + z\left({\cal A}_{\bz\to\frec} {\cal A}_{\bb\to\ftag} + {\cal A}_{\bb\to\frec} {\cal A}_{\bz\to\ftag}\right), \label{eq:etam}
\end{eqnarray}
\end{widetext}
where $\Gamma_d \equiv (\Gamma_H+\Gamma_L)/2$, $\Delta\Gamma_d \equiv \Gamma_H-\Gamma_L$,
$\dmd \equiv m_H-m_L$, $\Dt\equiv \trec - \ttag$, and
the ${\cal A}_{\bz,\bb\to\frec,\ftag}$ are the relevant decay amplitudes.
If $\frec$ is a $CP$ eigenstate ($\fcp$),
a parameter $\lambda_{CP}$, which characterizes $CP$ violation,
can be defined as $\lambda_{CP}\equiv (q/p) ({\cal A}_{\bb\to\fcp}/{\cal A}_{\bz\to\fcp})$.
The SM predicts $|\lcp|\simeq1$ and $\im(\etacplcp)\simeq\sin2\phi_1$ for the case $\fcp=\jpsi\kz$, where
$\eta_{CP}$ is the $CP$ eigenvalue of the final state.

\par

In this paper we report improved results on
the $CPT$-violating parameter $z$ and on the normalized total-decay-width difference $\dgog$ in
$\bz\to\jpsi\kz$ ($\kz=\ks,\kl$),
$\bz\to D^{(*)-}h^+$ ($h^+=\pip$ for $\Dm$ and $\pip, \rho^+$ for $D^{*-}$), and
$\bz \to D^{*-}\ell^+\nu_\ell$ ($\ell^+=e^+, \mu^+$) decays~\cite{bib:cc-disclaimer}.
Most of the sensitivity to $\rez$ and $\dgog$ is obtained from
neutral $B$-meson decays to $\fcp$,
while $\imz$ is constrained primarily from other neutral $B$-meson decay modes.

\par

The data sample of $535 \times 10^6$ $B\overline{B}$ pairs used in this analysis was collected with the Belle detector
at the KEKB asymmetric-energy $\elp\elm$ collider~\cite{bib:KEKB} (3.5 on 8.0~$\gev$)
operating at the $\ups$ resonance.
The $\ups$ is produced with a Lorentz boost of $\beta\gamma=0.425$ along the $Z$ axis,
which is antiparallel to the $e^+$ beam direction.
Since $B\overline{B}$ pairs  are produced approximately
at rest in the $\ups$ center-of-mass system (cms),
$\Dt$ can be approximated from $\Dz$, the difference between the $Z$ coordinates of the two $B$ decay vertices:\
$\Dt\simeq\Dz/(\beta\gamma c)$.

\par

The Belle detector~\cite{bib:Belle}
is a large-solid-angle magnetic spectrometer that consists of a silicon vertex detector (SVD),
a 50-layer central drift chamber, an array of aerogel Cherenkov counters,
a barrel-like arrangement of time-of-flight scintillation counters,
an electromagnetic calorimeter comprised of CsI(Tl) crystals,
located inside a superconducting solenoid coil that provides a 1.5~T magnetic field.
An iron flux-return located outside of the coil is
instrumented to detect $\kl$ mesons and to identify muons.
Two inner detector configurations are used;
a 2.0~cm radius beam pipe and a 3-layer SVD are used for the first data set (DS-I)
of $152\times10^6~B\overline{B}$ pairs,
while a 1.5~cm radius beam pipe, a 4-layer SVD, and a small-cell inner drift chamber are used to
record the remaining data set (DS-II) of $383\times10^6~B\overline{B}$ pairs.

\par

We reconstruct $\bz\to\frec$ decays in the
$\bz\to\jpsi\kz$, $\Dm\pip$, $D^{*-}\pip$, $D^{*-}\rho^+$, and $D^{*-}\ell^+\nu_\ell$ channels.
We also reconstruct $\bp\to\jpsi\kp$ and $\db\pip$
to precisely determine parameters for the
$\Dt$-resolution function model in neutral $B$ decays.
For the $\jpsi\ks$ and $\jpsi\kl$ modes,
we use the same selection criteria as in Ref.~\cite{bib:Belle-sin2phi1-535M}.
Candidate $\jpsi\kp$ events are selected from combinations of a charged track and a $\jpsi$ candidate
using the same selection criteria as in $\jpsi\ks$.
Charged and neutral charmed mesons are reconstructed in the $\Dm\to\kp\pim\pim$ and $\db\to\kp\pi^-, \kp\pi^-\pi^0, \kp\pi^-\pip\pi^-$ decay modes, respectively.
The invariant mass of their daughters, $M_{Kn\pi}$,
is required to be within $45~\mevcc$~($\sim\!\!5\sigma$) of the nominal $D$-meson mass for the mode with $\pi^0$,
or $30~\mevcc$~($\sim\!\!6\sigma$) for the other modes.
Candidate $D^{*-}$ mesons are reconstructed in $\db\pim$ combinations,
in which the mass difference $M_{\rm diff}$
between the $D^{*-}$ and $\db$ candidates is required to be within $5~\mevcc$~($\sim\!\!8\sigma$) of the nominal value.
Candidate $\rho^+$ mesons are reconstructed from $\pip\pi^0$ combinations with invariant mass
within $225~\mevcc$ of the nominal $\rho^+$ mass.
The $D^{*-}$ candidates for the final state $D^{*-}\ell^+\nu_\ell$ are reconstructed using the $D^{*-}$ and $\db$ decay modes listed above, where
the detailed selection criteria are described in Ref.~\cite{bib:Belle-sin2phi1-152M}.

\par

We identify $\bz$ or $\bp$ candidates in modes
other than $\bz\to\jpsi\kl$ or $D^{*-}\ell^+\nu_\ell$ using the beam-energy constrained mass,
$M_{\rm bc}\equiv\sqrt{(E_{\rm beam}^*)^2-|\vec{p}_{B}^*|^2}$,
and the energy difference, $\Delta E \equiv E_B^* - E_{\rm beam}^*$,
where $E_{\rm beam}^*$ is the beam energy in the cms, and $E_B^*$ and
$\vec{p}_{B}^*$ are the cms energy and momentum of
the reconstructed $B$ candidate, respectively.
The signal region for the $M_{\rm bc}$ is defined as $5.27~\gevcc < M_{\rm bc} < 5.29~\gevcc$
for all decay modes, while
that for $\Delta E$ is decay-mode dependent:
$|\Delta E| < 40~\mev$ for $\jpsi\ks$ and $\jpsi\kp$;
$|\Delta E| < 45~\mev$ for $\Dm\pip$;
$|\Delta E| < 70~\mev$ for $D^{*-}\pip$;
$-50~\mev < \Delta E < +80~\mev$ for $D^{*-}\rho^+$, and
$|\Delta E| < 60~\mev$ for $\db\pip$.
Candidate $\bz\to\jpsi\kl$ decays are selected
by requiring $0.20~\gevc < |\vec{p}_{B}^*| < 0.45~\gevc$.
For $\bz\to D^{*-}\ell^+\nu_\ell$ decays,
the energies and momenta of the $B$ meson and $D^*\ell$ system in the cms satisfy
$M_{\nu_\ell}^2 = (E_B^* - E_{D^{*-}\ell^+})^2 - (|\vec{p}_{B}^*|^2 + |\vec{p}_{D^*\ell}^*|^2 - 2 |\vec{p}_{B}^*|\,|\vec{p}_{D^*\ell}^*|\cos\theta_{B,D^*\ell})$,
where $M_{\nu_\ell}$ is the neutrino mass and
$\cos\theta_{B,D^*\ell}$ is the angle between $\vec{p}_{B}^*$
and $\vec{p}_{D^*\ell}^*$.  We calculate $\cos\theta_{B,D^*\ell}$ setting $M_{\nu_\ell}=0$ and
$E_B^* = E_{\rm beam}^*$.
The signal region is defined as $|\cos\theta_{B,D^*\ell}| < 1.1$.
In the $\cos\theta_{B,D^*\ell}$ signal region, $\bz\to D^{**-}\ell^+\nu_\ell$ decays are also reconstructed.
Since the $\Dt$ distribution is expected
to be the same as that in $D^{*-}\ell^+\nu_\ell$, we treat $\bz\to D^{**-}\ell^+\nu_\ell$ decays as signal.

\par

The event-by-event signal and background probabilities are estimated
from signal and background distributions of the kinematic parameters,
$M_{\rm bc}$, $\Delta E$, $|\vec{p}_{B}^*|$, and $\cos\theta_{B,D^*\ell}$.
Signal and combinatorial background distributions in $M_{\rm bc}$ are modeled by
Gaussians and an empirically determined background shape
with a kinematic threshold originally introduced by ARGUS~\cite{bib:ARGUS}, respectively,
while those in $\Delta E$ are modeled by the sum of two Gaussians
and a first-order polynomial, respectively.
The model parameters for the signal and combinatorial background distributions
in the $\jpsi\ks$ and $\jpsi\kp$ modes are determined
from a two-dimensional fit to the $M_{\rm bc}$-$\Delta E$ distributions in data.
In Monte Carlo (MC) simulations of the $D^{(*)-}h^+$ and $\db\pip$ modes,
in addition to combinatorial background, we find a background contribution, which comes from
charged and neutral $B$-meson decays with one or more particles missed in their reconstruction,
and which peaks in $M_{\rm bc}$ (peaking background).
The model parameters for the signal and combinatorial background distributions
are determined from signal and sideband $M_{\rm bc}$ distributions in data,
while those of the peaking background are modeled by an ad-hoc distribution obtained from MC simulation.
In addition to the combinatorial background,
we find from MC simulation that the background in
$\jpsi\kl$ is mainly comprised of $(c\bar{c})\kz$ modes
except for contributions from $\jpsi\kl$, $\jpsi\kz\pi^0$, $\jpsi\pi^0$, and charged $B$-meson decays.
For $D^{*-}\ell^+\nu_\ell$, there is an additional background from $\overline{D}^{**0}\ell^+\nu_\ell$.
For the $\jpsi\kl$ and $D^{*-}\ell^+\nu_\ell$ modes,
the signal and noncombinatorial background distributions
in $|\vec{p}_{B}^*|$ and $\cos\theta_{B,D^*\ell}$, respectively,
are modeled using MC simulation, while the combinatorial background distributions are obtained from
sideband regions of the $\jpsi$ and $D^{*-}$, respectively.

\par

The $b$-flavor of $\ftag$ is identified from inclusive
properties of particles that are not associated with the $\bz,\bb\to\frec$ decay.
The tagging information is represented by two event-by-event parameters,
the $b$-flavor charge $\qtag$ and an MC-determined flavor-tagging dilution factor $r$~\cite{bib:flavor-tag}.
The parameter $r$ ranges from $r=0$ for no flavor discrimination to $r=1$ for unambiguous flavor assignment.
For events with $r>0.1$, the wrong tag fractions for six $r$ intervals, $w_l~(l=1\ldots6)$,
and their differences between $\bz$ and $\bb$ decays, $\Delta w_l$,
are determined using the data sample as described later.
If $r\le0.1$, we set the wrong tag fraction to 0.5 so that the event is not used on flavor tagging.

\par

The vertex position is reconstructed using charged tracks that have sufficient SVD hits~\cite{bib:resolution}.
The $\frec$ vertex for the modes with a $\jpsi$ is reconstructed using lepton tracks from the $\jpsi$ decay,
while in modes without a $\jpsi$ the $\frec$ vertex is reconstructed by combining
the $\db$- or $\Dm$-meson trajectory and the remaining charged track forming the $B$-meson candidate;
the slow $\pi^-$ from the $D^{*-}$ decay is not included because of its poor position resolution.
The $\ftag$ vertex is obtained from selected well-reconstructed tracks that are not assigned to $\frec$.
A constraint on the interaction region profile (IP) in the plane perpendicular to the $Z$ axis
is also applied to both $\frec$ and $\ftag$ reconstructed vertices.
We model the resolution function $R(\Dt)$ as a convolution of four sub-components~\cite{bib:resolution}:
detector resolutions for $\frec$ and $\ftag$ vertex reconstruction,
boost effect due to nonprimary particle decays in $\ftag$, and
dilution by the kinematic approximation $\Dt\simeq\Dz/(\beta\gamma c)$.
Nearly all model parameters are determined using the data as described later.
The exceptions are the parameters for the boost effect and kinematic
approximation, which are obtained using MC simulation.
For candidate events in which both $B$ vertices are found,
for further analysis, we only use events with vertices that satisfy
$\xi_{\rm rec}<250$, $\xi_{\rm tag}<250$, and $|\Dt|<70$~ps,
where $\xi_{\rm rec}$ ($\xi_{\rm tag}$) is the $\chi^2$
of the $\frec$ ($\ftag$) vertex fit calculated only along the $Z$ direction~\cite{bib:Belle-sin2phi1-152M}.

\par

After flavor tagging and vertex reconstruction,
we count the number of events remaining in
the signal region $N_{\rm ev}$ and estimate the purity for each decay mode.
The values of $N_{\rm ev}$ and purity for each mode are listed in Table~\ref{tab:nev}.
\begin{table}
\caption{Number of events  $N_{\rm ev}$ and purity in the signal region for each decay mode.}
\label{tab:nev}
\begin{ruledtabular}
\begin{tabular}{lcc}
$B$ decay mode   & $N_{\rm ev}$ & Purity (\%) \\
\hline
$\jpsi\ks$             & $7713$   & $97.0$  \\
$\jpsi\kl$             & $10966$  & $59.2$  \\
$\Dm\pip$              & $39366$  & $83.2$  \\
$D^{*-}\pip$           & $46292$  & $81.5$  \\
$D^{*-}\rho^+$         & $45913$  & $66.3$  \\
$D^{*-}\ell^+\nu_\ell$ & $383818$ & $75.2$  \\
$\jpsi\kp$             & $32150$  & $97.3$  \\
$\db\pip$              & $216605$ & $63.9$  \\
\end{tabular}
\end{ruledtabular}
\end{table}

\par

We determine three major physics parameters $\rez$, $\imz$, and $\dgog$
together with five other physics parameters $\taubz$, $\taubp$ (neutral and charged $B$-meson lifetimes),
$\dmd$, $|\lcp|$, and $\arg(\etacplcp)$
in a simultaneous 72-parameter fit to the observed $\Dt$ distribution.
The remaining 64 parameters are the
$\Dt$-resolution function model parameters (34),
flavor-tagging parameters $w_l$ and $\Delta w_l$ (24), and
background parameters for $\bz\to D^{*-}\ell^+\nu_\ell$ (6).
The nonphysics parameters are determined separately for DS-I and DS-II.
An unbinned fit is performed by maximizing a likelihood function defined by
$L(\rez,\imz,\dgog) = \prod_{i} L^i(\rez,\imz,\dgog;\Dt^i,q_{\rm tag}^i)$,
where the product is over all events in the signal region.
The likelihood for the $i$-th event $L^i$ is given by
\begin{eqnarray}
\lefteqn{L^i = (1-f_{\rm ol} ) f_{\rm sig}^i {\cal P}(\Dt^i;f_{\rm rec}^i,f_{\rm tag}^i) \otimes R^i(\Dt^i)} && \nonumber \\
&& \quad + (1-f_{\rm ol} ) \sum_k f^{k,i}_{\rm bkg} P^k_{\rm bkg}(\Dt^i) + f_{\rm ol} P_{\rm ol}(\Delta t^i).
\label{eq:likelihood}
\end{eqnarray}
The first term accounts for the signal component,
where $f^i_{\rm sig}$ is an event-by-event signal fraction.
In Eq.~(\ref{eq:likelihood}) ${\cal P}$ is modified from Eq.~(\ref{eq:master_PDF}) by including the event-by-event
incorrect-tagging effect, $w_l^i$ and $\Delta w_l^i$, and
the symbol $\otimes$ indicates a convolution with the $\Dt$-resolution function $R^i(\Dt)$.
The second term accounts for the background component,
where $f^{k,i}_{\rm bkg}$ is an event-by-event background fraction and $k$ runs over all background components.
The signal and background fractions are normalized to $f^i_{\rm sig}+\sum_k f_{\rm bkg}^{k,i} = 1$.
The $\Dt$-distribution for the combinatorial background component is modeled
using the sideband region of $\Delta E$-$M_{\rm bc}$, $|\vec{p}_{B}^*|$, or $\cos\theta_{B,D^*\ell}$ space,
while the $\Dt$ distribution for the peaking-background components are modeled by MC simulation.
The third term accounts for a small but broad
($\Dt$ outlier) component that cannot be described by the first and second terms,
where $f_{\rm ol}$ is an event-dependent outlier fraction and $P_{\rm ol}(\Dt)$ is a broad
Gaussian.
In the nominal fit, we account for
Cabibbo-Kobayashi-Maskawa-favored $B \to \overline{D}$ decay via $b\to c\bar{u}d$ (CFD) but neglect the contribution
from Cabibbo-Kobayashi-Maskawa-suppressed $B \to D$ decay via $b\to u \bar{c}d$ (CSD)
both in $\frec$ and $\ftag$.
The effect of the CSD is included in the systematic uncertainty.

\par

\begin{figure}[ht]
\begin{tabular}{ll}
\includegraphics[width=0.240\textwidth]{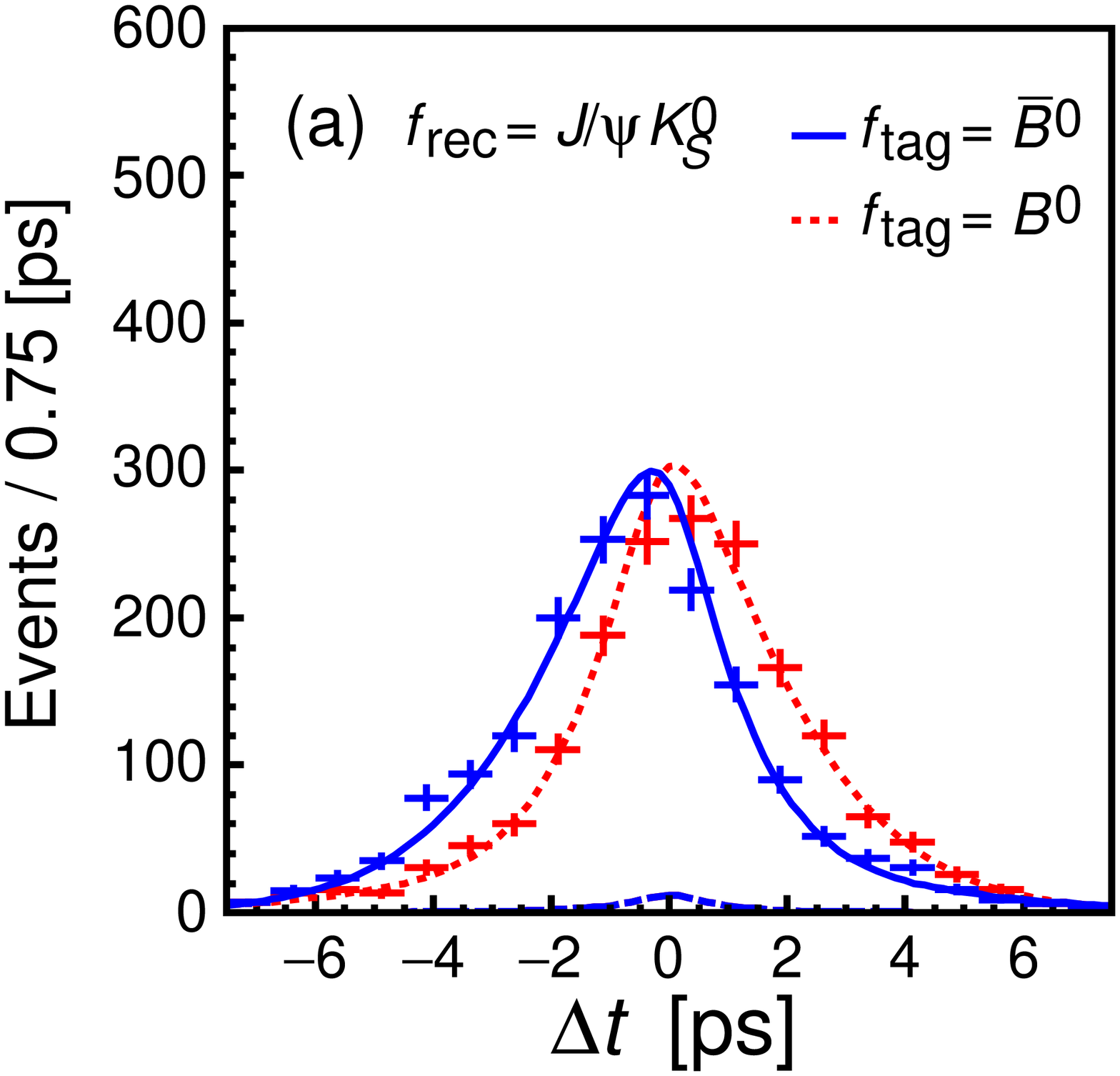} &
\includegraphics[width=0.240\textwidth]{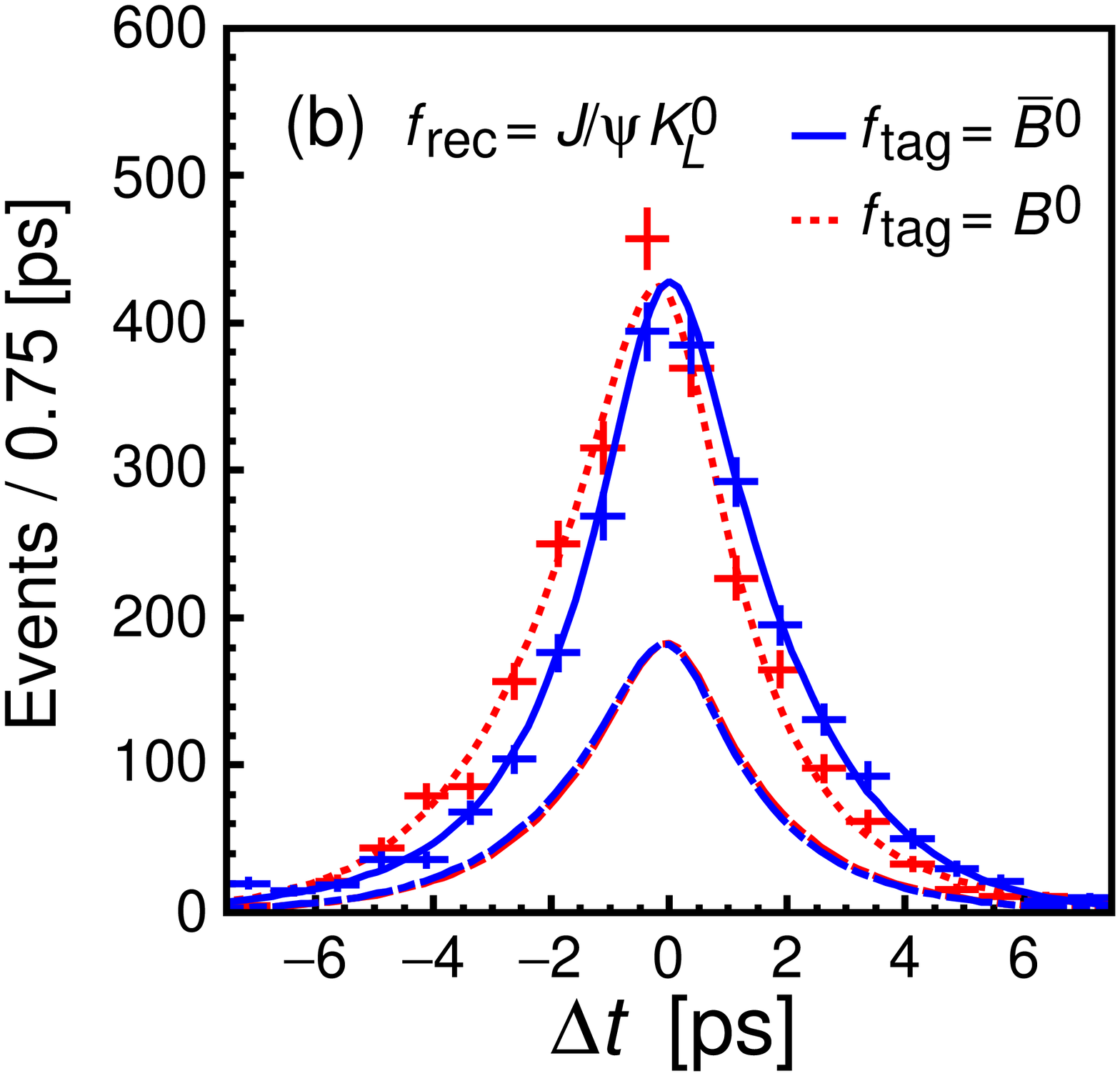} \\
\end{tabular}
\caption{
$\Dt$ distributions for events with flavor tag quality $r>0.5$,
where (a) and (b) correspond to $\frec=\jpsi\ks$ and $\jpsi\kl$ cases, respectively.
Events are separated according to tagged $\ftag$ flavor,
where the solid and dashed curves are for $\qtag=+1$ and $-1$ events, respectively.
The two chain curves below the fit curves indicate the sum of the background and $\Dt$-outlier components
for each $\ftag$ flavor, which are almost indistinguishable because of their similar shapes.
\label{fig:cptfit_dt_cp}
}
\end{figure}

\begin{figure}[ht]
\begin{tabular}{ll}
\includegraphics[width=0.240\textwidth]{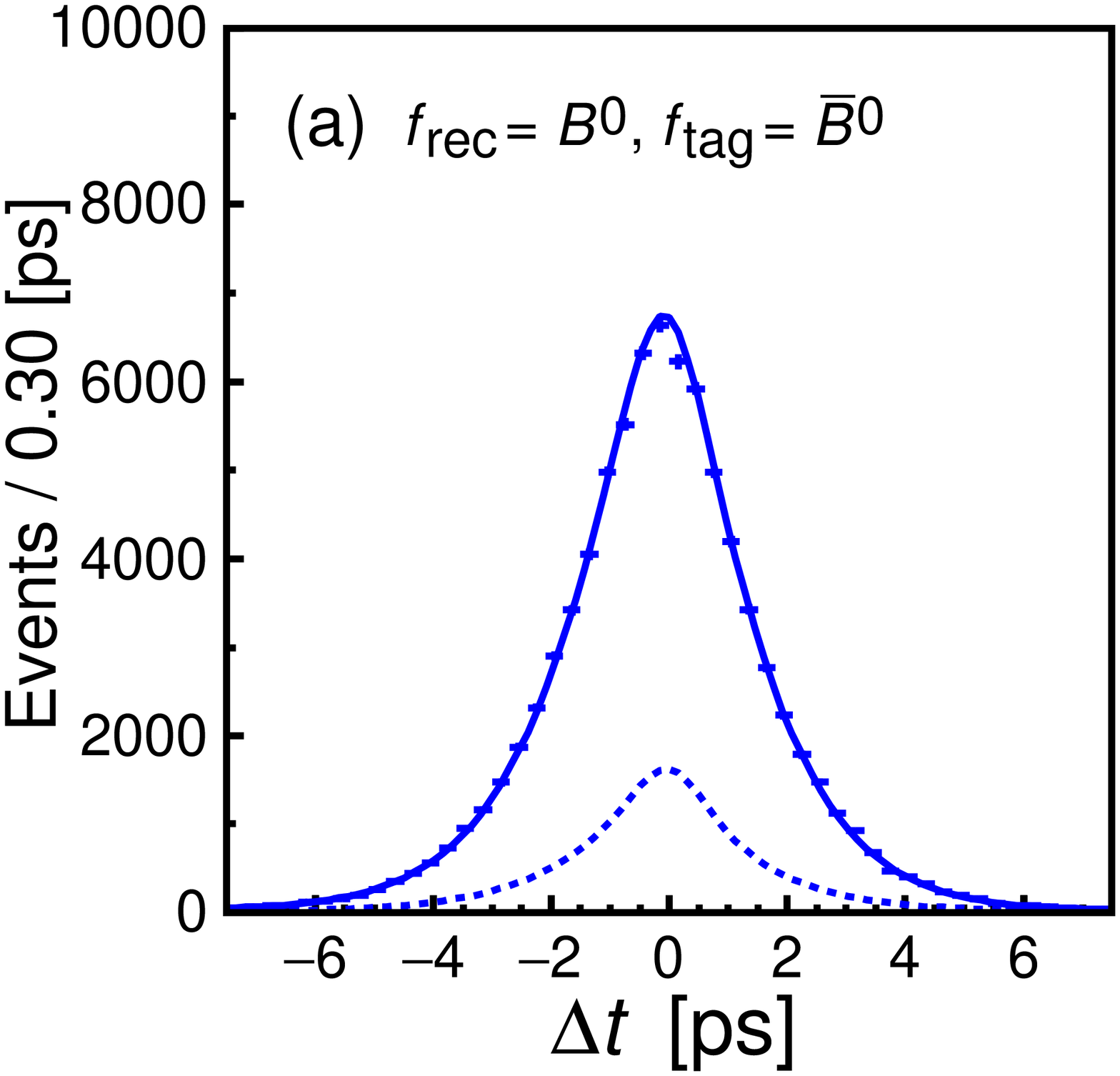} &
\includegraphics[width=0.240\textwidth]{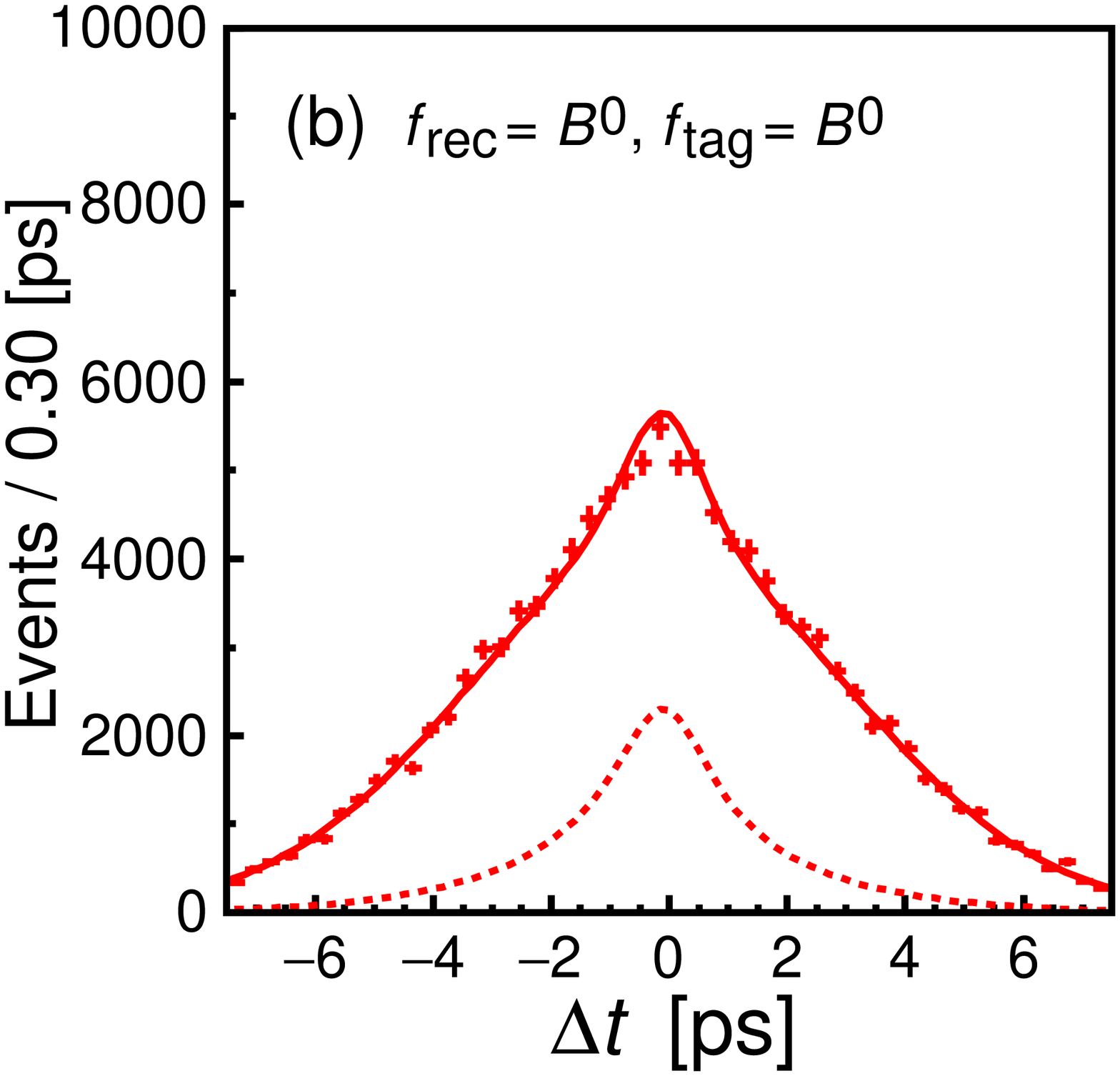} \\
\includegraphics[width=0.240\textwidth]{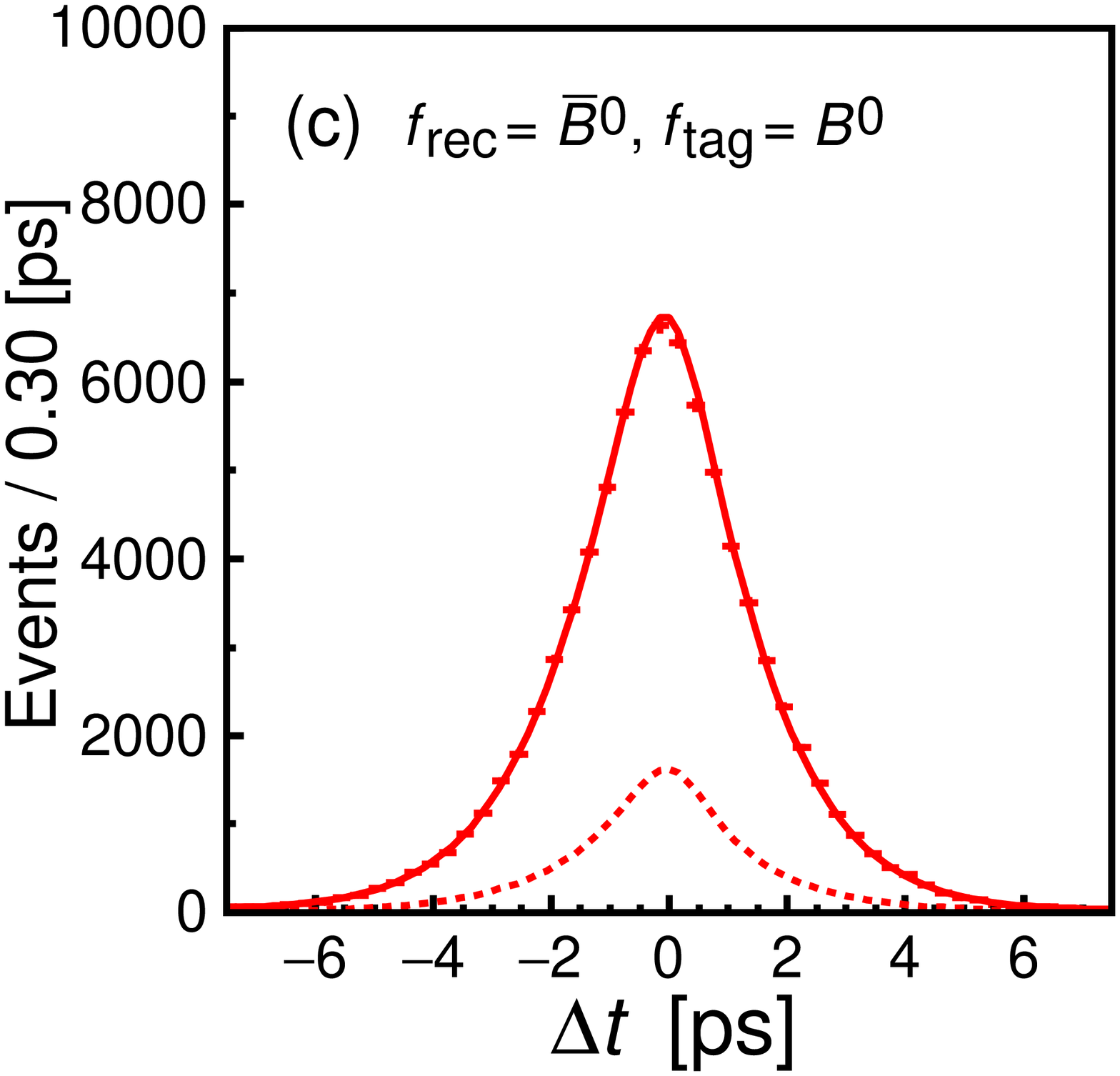} &
\includegraphics[width=0.240\textwidth]{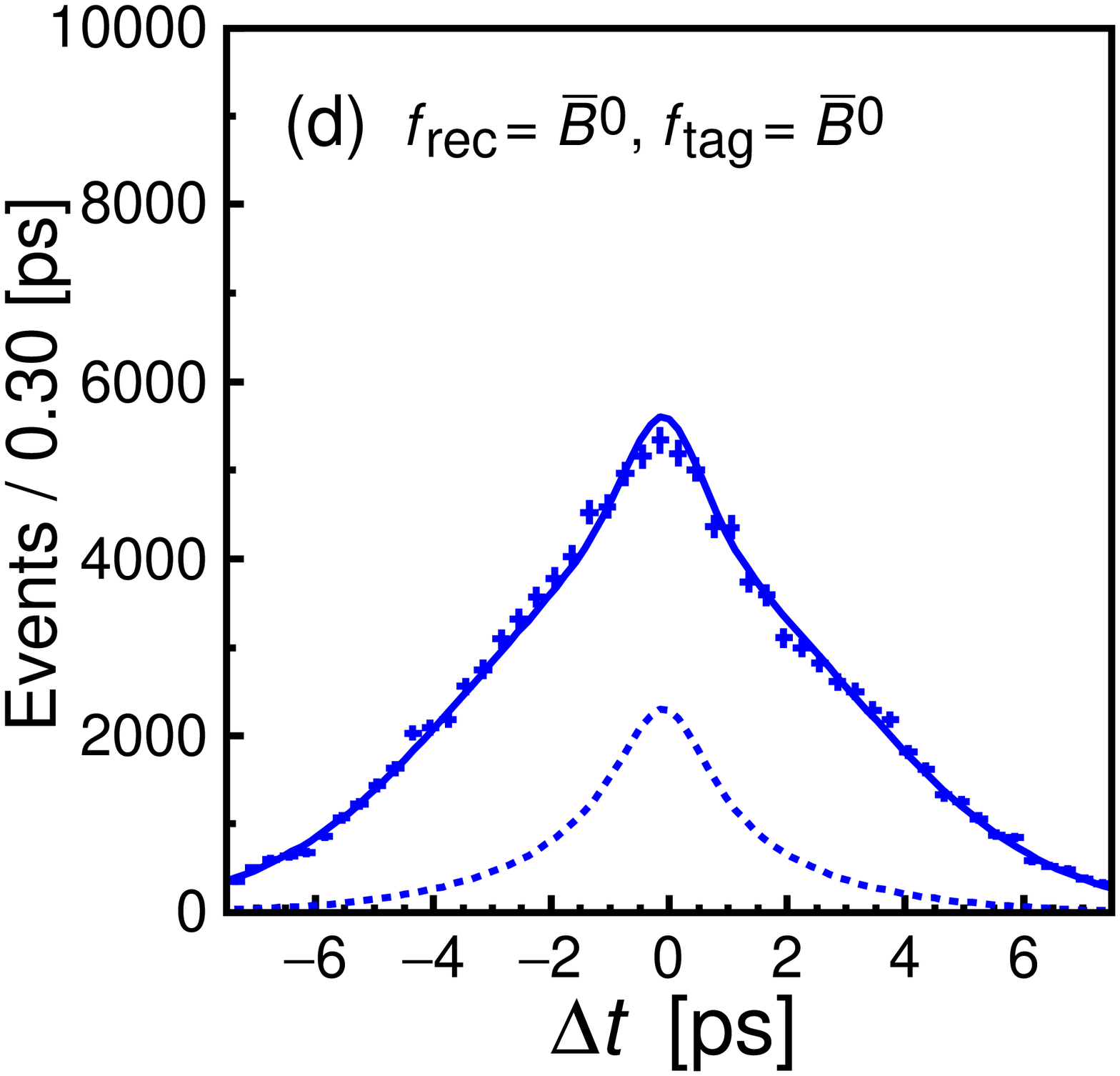} \\
\end{tabular}
\caption{
$\Dt$ distributions for events with flavor tag quality $r>0.5$,
where (a, b) and (c, d) correspond to flavor-specific $\frec=\bz$ and $\bb$ cases, respectively.
The dashed curve below the solid fit curve is the sum of the background and $\Dt$-outlier components.
\label{fig:cptfit_dt_nocp}
}
\end{figure}

From the fit to the data, we obtain
$\rez=(+1.9\pm3.7)\times10^{-2}$,
$\imz=(-5.7\pm3.3)\times10^{-3}$, and
$\dgog=(-1.7\pm1.8)\times10^{-2}$, together with
$\taubz=1.531\pm0.004$~ps,
$\taubp=1.640\pm0.006$~ps,
$\dmd=0.506\pm0.003$~ps$^{-1}$,
$|\lcp|-1 = (1.1\pm3.8)\times10^{-3}$, and
$\arg(\etacplcp) = -0.700\pm0.042$,
where all uncertainties are statistical only.
The fit has a twofold ambiguity in the sign of $\re(\etacplcp)$;
$\rez$ and $\dgog$ change signs depending on its sign.
We take the solution with positive $\re(\etacplcp)$, which is the
result of the global fit ~\cite{bib:CKMfitter-ICHEP2010}.
The correlation coefficients $\rho$ between two of the three major physics parameters are
$\rho_{\rez,\imz}= -0.17$, $\rho_{\rez,\dgog}=+0.08$, and $\rho_{\imz,\dgog}=+0.09$.
The largest correlation coefficient between a major physics parameter
and any other fit parameter is $\rho_{\rez,\dmd}=+0.24$.
The fitted values of $|\lcp|$ and $\arg(\etacplcp)$ give $\sin2\phi_1 = 0.645\pm0.032\mbox{(stat)}$,
which is consistent with our dedicated $\sin2\phi_1$ measurement
with the same data sample~\cite{bib:Belle-sin2phi1-535M},
because the major physics parameters are consistent with zero.
Figures~\ref{fig:cptfit_dt_cp} and~\ref{fig:cptfit_dt_nocp} show the $\Dt$ distributions
for events with $\frec=\jpsi\kz$ cases and the other cases, respectively,
with the fitted curves superimposed.

\begin{figure}[ht]
\begin{tabular}{ll}
\includegraphics[width=0.240\textwidth]{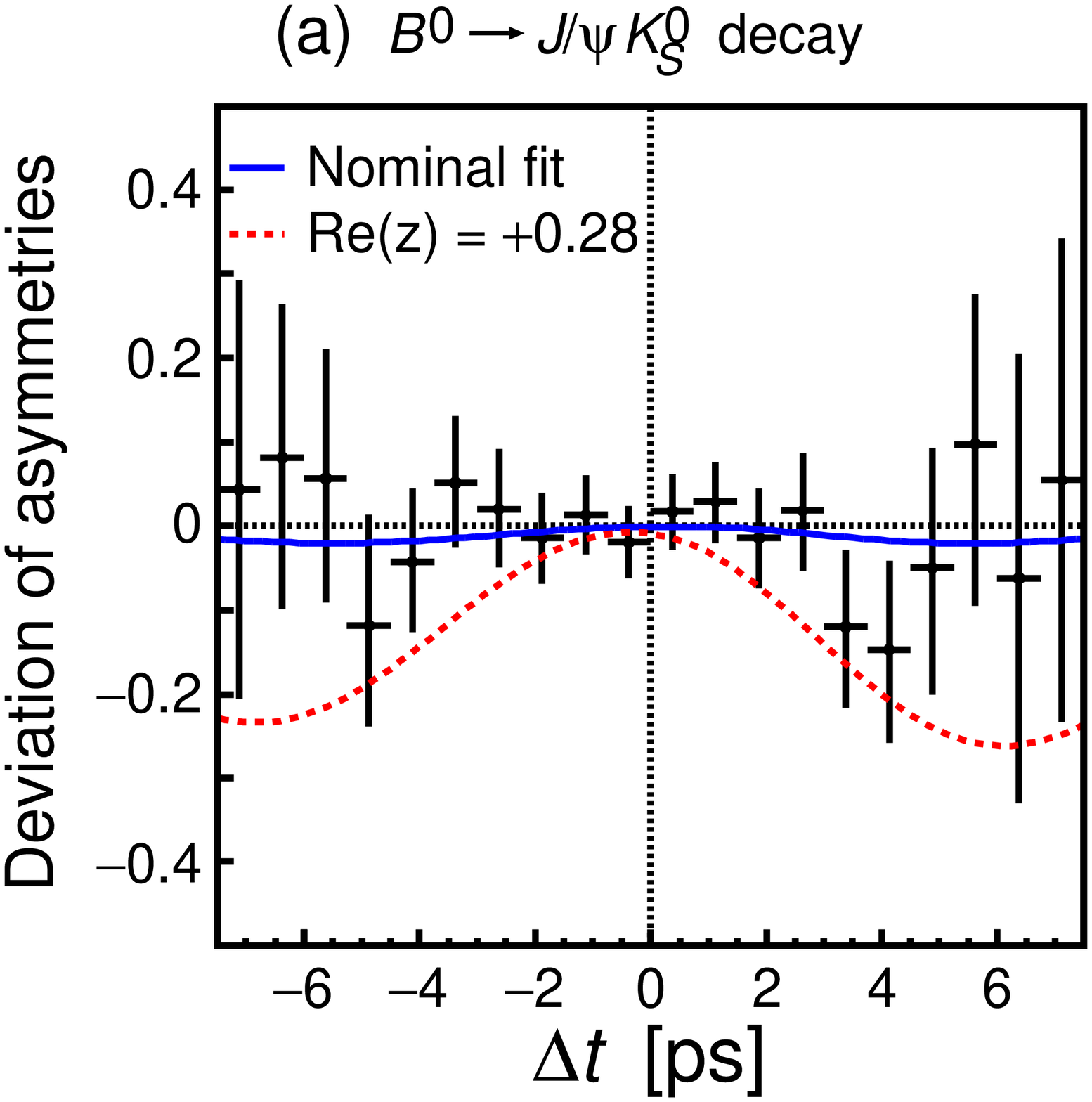} &
\includegraphics[width=0.240\textwidth]{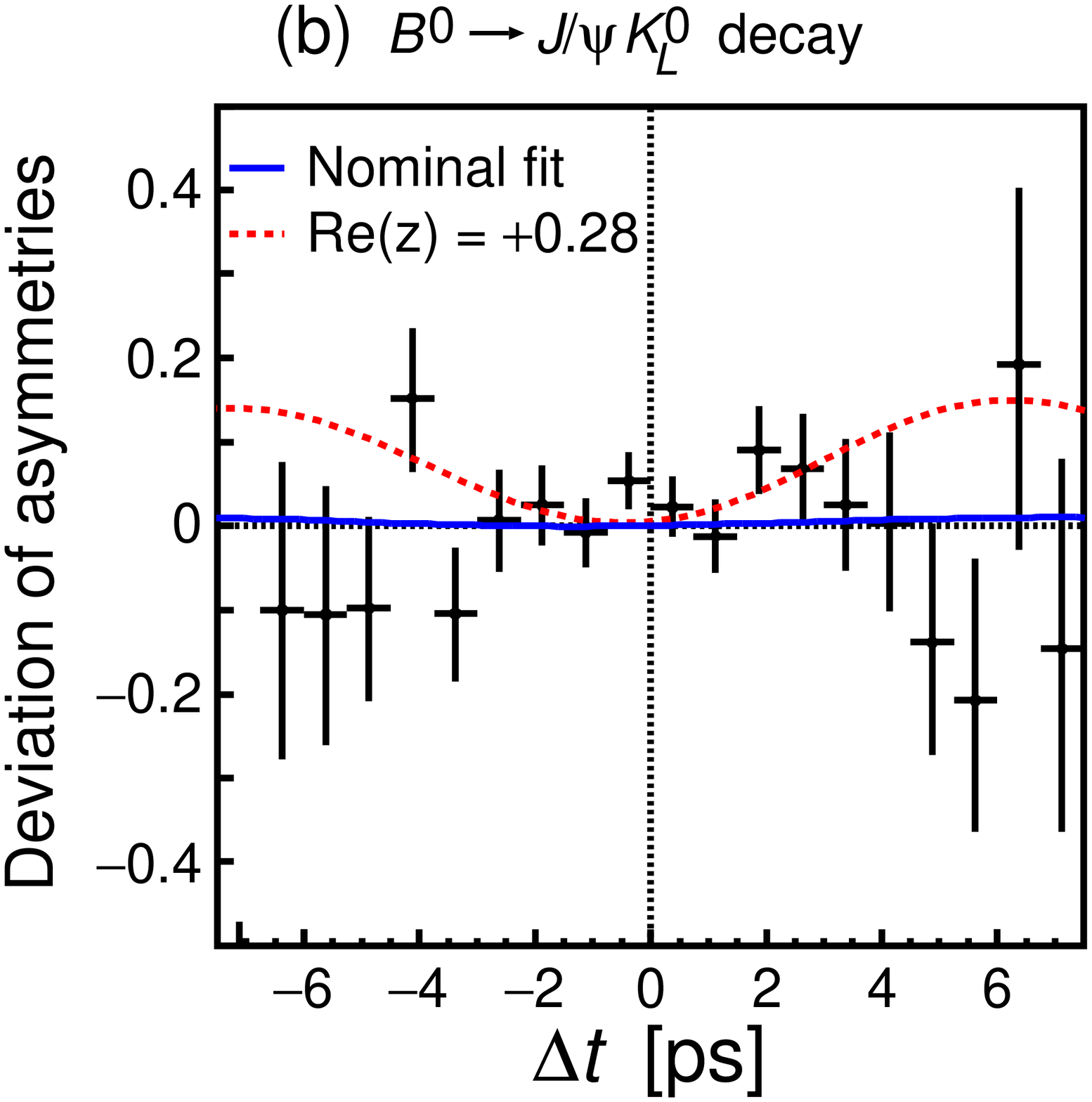} \\
\\
\includegraphics[width=0.240\textwidth]{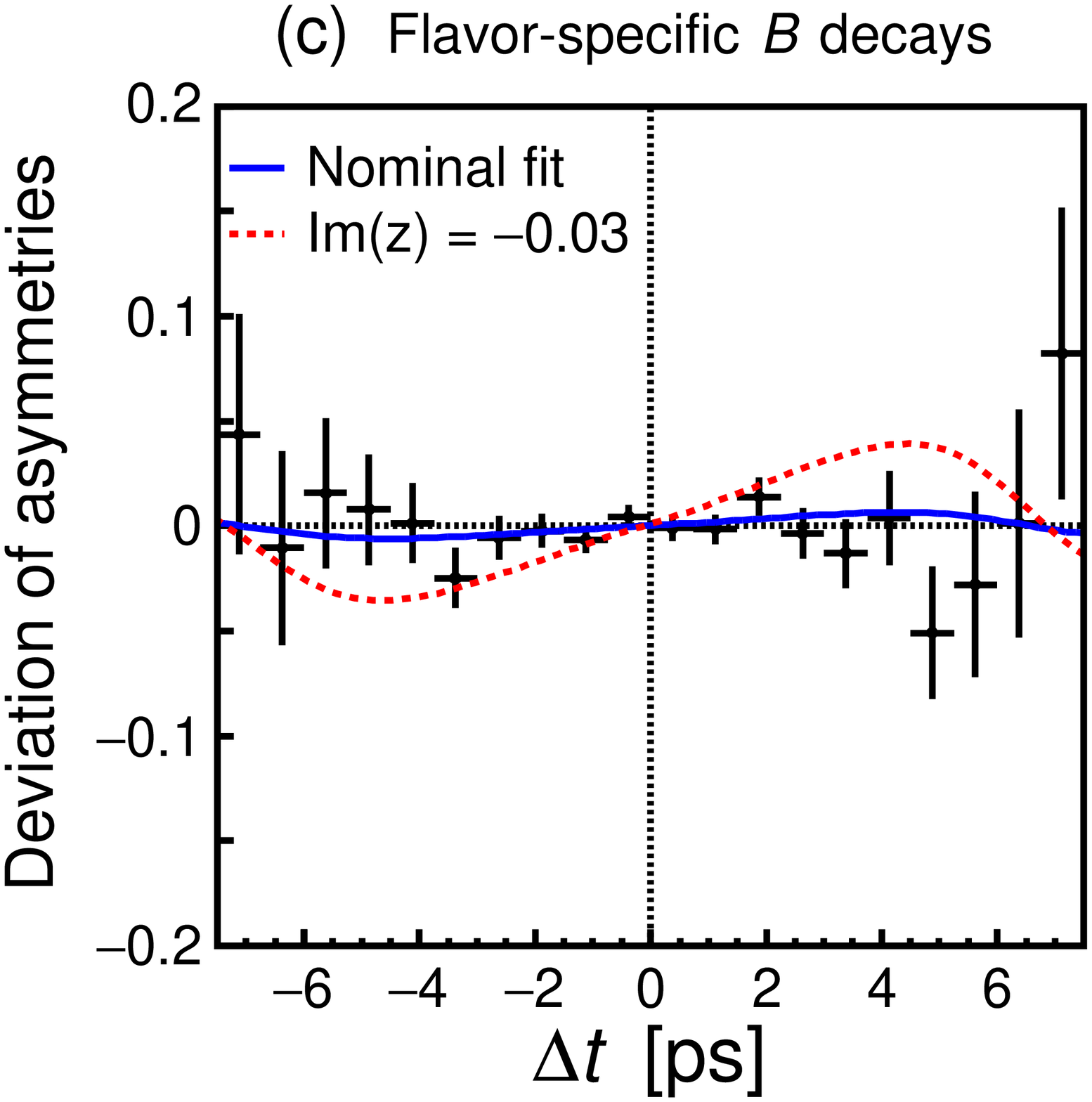} &
\includegraphics[width=0.240\textwidth]{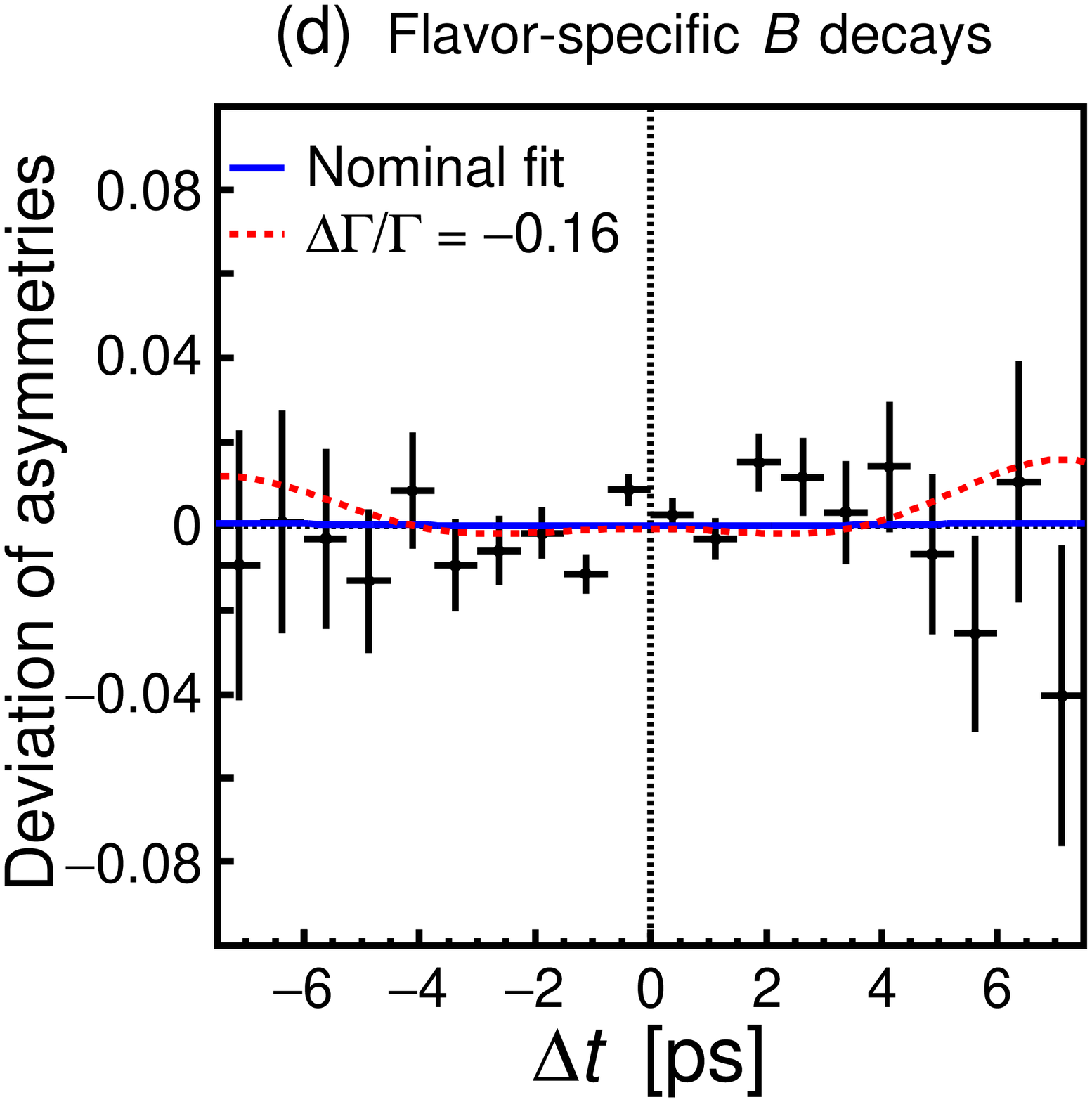} \\
\end{tabular}
\caption{
Deviations of the asymmetries from the reference asymmetry.
The crosses with error bars are data.  The solid curves are deviations for the nominal fits.
The dashed curves are with $\rez=+0.28$ for (a) and (b),
$\imz=-0.03$ for (c), and $\dgog=-0.16$ for (d) (see text for details).
\label{fig:cptfit_demo}
}
\end{figure}

To illustrate the $CPT$ sensitivity of our measurements, we plot the deviations of the
asymmetries from a reference asymmetry obtained from the nominal fit parameters but setting $\rez=\imz=\dgog=0$ in Fig.~\ref{fig:cptfit_demo},
where (a), (b), and (c) show those for $CP$ asymmetries of $\bz\to\jpsi\ks$,
$\jpsi\kl$, and opposite-flavor $B$-meson pairs, respectively;
(d) shows asymmetries between the opposite-flavor and same-flavor
$B$-meson pairs.  Asymmetries are obtained from events in $\Dt$ bins
without background subtraction, where the events are required
to have $r>0.5$.
We superimpose the deviations of the asymmetries
for the nominal fit curves and those with one parameter shifted
by $\sim5$ times the statistical uncertainty in each subsample fit.
For illustration, the most appropriate parameter is chosen in each plot.

\par

Table~\ref{tab:systematic-error} lists the systematic uncertainties on the major physics parameters.
The total systematic uncertainty is obtained by adding the contributions in Table~\ref{tab:systematic-error} in quadrature.
The dominant contributions
are from the tag-side interference (TSI)~\cite{bib:TSI}
and vertex reconstruction; the next largest contributions are from fit bias.

\begin{table}
\caption{Summary of systematic uncertainties on the major physics parameters.}
\label{tab:systematic-error}
\begin{ruledtabular}
\begin{tabular}{lrrr}
Source                 & $\delta(\rez)$  & $\delta(\imz)$ & $\delta(\dgog)$ \\
\hline
Vertex reconstruction    & 0.008             & 0.0028            & 0.009 \\
$\Dt$-resolution function& 0.003             & 0.0004            & 0.002 \\
Tag-side interference    & 0.028             & 0.0006            & 0.001 \\
CSD effect               & 0.004             & 0.0008            & 0.003 \\
Fit bias                 & 0.012             & 0.0013            & 0.005 \\
Signal fraction          & 0.004             & 0.0002            & 0.002 \\
Background $\Dt$ shape   & 0.005             & 0.0001            & 0.002 \\
Others                   & 0.001             & $<0.0001$         & 0.002 \\
\hline
Total                  & 0.033             & 0.0033            & 0.011 \\
\end{tabular}
\end{ruledtabular}
\end{table}

\par

The TSI effect arises from the interference between
CFD and CSD amplitudes in $\ftag$.
In general, the presence of CSD introduces new terms
in Eqs.~(\ref{eq:etap})~and~(\ref{eq:etam})
\begin{eqnarray}
{\cal A}_{\bz\to {f_{\rm CSD}}} &=& R_\frec \exp \left[\,i\, (+\phi_3+\delta_\frec) \right], \nonumber \\
{\cal A}_{\bb\to {f_{\rm CSD}}} &=& R_\frec \exp \left[\,i\, (-\phi_3+\delta_\frec) \right], \label{eq:CSD_amplitudes}
\end{eqnarray}
where $R_\frec$ and $\delta_\frec$ are
the mode-dependent ratio of the CSD to CFD amplitudes and
the relative strong-phase difference between the CSD and CFD amplitudes, respectively,
and $\phi_3=67.2^\circ$~\cite{bib:CKMfitter-ICHEP2010}.
For the tag-side parameter, $R_\ftag$ and $\delta_\ftag$ are ``effective'' values
because $\ftag$ is an admixture of several decay modes, some of which do not have a corresponding CSD.
The effective $R_\ftag$ and $\delta_\ftag$ parameters are estimated using
the $\bz \to D^{*-}\ell^+\nu_\ell$ sample~\cite{bib:Belle-sin2phi1-152M}.
We perform fits to the major physics parameters varying the terms from Eqs.~(\ref{eq:CSD_amplitudes})
into Eqs.~(\ref{eq:etap})~and~(\ref{eq:etam}).
The deviation from the nominal fit is quoted as a systematic uncertainty.

\par

The CSD effects in $\frec$ are investigated by performing fits of the major physics parameters
varying the $R_\frec$ and $\delta_\frec$ parameters introduced in Eqs.~(\ref{eq:CSD_amplitudes}).
For the $\Dm\pip$ and $D^{*-}\pip$ modes,
we use $R_{D\pi}=0.02$ or $R_{D^*\pi}=0.02$ predicted in Ref.~\cite{bib:R-prediction},
and $\delta_{D^{(*)}h}$ computed from measurements of $CP$-violating parameters
in the relevant $B$ decays~\cite{bib:phi3}.
We quote fitted deviations as the systematic uncertainties.
For the $D^{*-}\rho^+$ mode,
we assume $R_{D^*\rho}=0.02$,
and allow $\delta_{D^*\rho}$ to be $0^\circ$, $90^\circ$, $180^\circ$, or $270^\circ$,
because of the absence of $CP$-violating parameter measurements.
We quote the largest fitted deviation as the systematic uncertainty.

\par

The systematic uncertainty due to vertex reconstruction is estimated
as follows.  We repeat fits by changing various requirements or
parameters used in the vertex reconstruction: the IP constraint,
the track selection criteria,
and the calibration of the track position and momentum uncertainties.
The deviation from the nominal fit is quoted as the systematic uncertainty.
Systematic errors due to imperfect SVD alignment are estimated from MC samples
with artificially varied alignment constants.
Effects from small biases in the $\Dz$ measurement observed in $e^+e^- \to \mu^+\mu^-$
and other control samples are accounted for by applying a special correction function
and including the variation from the nominal result into the systematic uncertainty.

\par

We estimate the fit biases $\delta^{\rm bias}_\rez$, $\delta^{\rm bias}_\imz$, and $\delta^{\rm bias}_\dgog$
using an analysis procedure with fully simulated MC samples.
We generate sets of $\Dt$ distributions with statistics similar to data,
fixing $(\rez,\imz,\dgog)=(0,0,0)$ or
varying one of the three input parameters to $\rez=\pm0.01$, $\imz=\pm0.01$, or $\dgog=\pm0.05$.
We perform a full-parameter fit to each generated distribution without the background component,
and take deviations of the fitted three parameters from the input value as the bias.
We quote the average value of biases in the above seven samples.
These effects are included into the systematic uncertainty after symmetrization.

\par

The systematic uncertainty due to the $\Dt$-resolution function is estimated
by varying by $\pm2\sigma$ each resolution-function parameter determined from MC,
and repeating the fit to add each variation in quadrature.
We also take the systematic effect from the $\Dt$-outlier elimination criteria
into account in the systematic uncertainty by varying each criterion
and adding each variation in quadrature.

\par

The most precise previous results on the $CPT$-violating parameter and $\dgog$
in the neutral $B$-meson system were obtained by the {\it BaBar} collaboration.  They found
${\cal R}e(\lcp/|\lcp|)\rez = +0.014\pm0.035\mbox{(stat)}\pm0.034\mbox{(syst)}$,
$\imz = (-13.9\pm7.3\mbox{(stat)}\pm3.3\mbox{(syst)})\times10^{-3}$, and
${\rm sgn}(\re(\lcp))\dgog = -0.008\pm0.037\mbox{(stat)}\pm0.018\mbox{(syst)}$~[\onlinecite{bib:BaBar-CPT-Hadronic},\onlinecite{bib:BaBar-CPT-Dilepton}].
For ${\cal R}e(\lcp/|\lcp|)\rez$, our result is $(+1.5\pm3.8)\times10^{-2}$, where the total error is quoted.
Our result is consistent with Ref.~\cite{bib:BaBar-CPT-Dilepton}
and improves the overall precision
by factors of 1.3 to 2.0 for all parameters.

\par

In summary, we report a new search for $CPT$ violation with an
improved measurement of the $CPT$-violating parameter $z$ and
normalized decay-rate difference $\dgog$
in $\bz\to \jpsi\ks$, $\jpsi \kl$, $\Dm\pip$, $D^{*-}\pip$, $D^{*-}\rho^+$, and
$D^{*-}\ell^+\nu_\ell$ decays using $535 \times 10^6$ $B\overline{B}$ pairs collected at the
$\ups$ resonance with the Belle detector.
We find
\begin{eqnarray*}
\rez &=&[+1.9\pm3.7\mbox{(stat)}\pm3.3\mbox{(syst)}]\times10^{-2},\\
\imz &=&[-5.7\pm3.3\mbox{(stat)}\pm3.3\mbox{(syst)}]\times10^{-3}, \mbox{ and}\\
\dgog&=&[-1.7\pm1.8\mbox{(stat)}\pm1.1\mbox{(syst)}]\times10^{-2},
\end{eqnarray*}
all of which are consistent with zero.
This is the most precise measurement of $CPT$-violating parameters in the neutral $B$-meson system to date.

\par

We thank the KEKB group for excellent operation of the
accelerator; the KEK cryogenics group for efficient solenoid
operations; and the KEK computer group, the NII, and
PNNL/EMSL for valuable computing and SINET4 network support.
We acknowledge support from MEXT, JSPS and Nagoya's TLPRC (Japan);
ARC and DIISR (Australia); NSFC (China); MSMT (Czechia);
DST (India); INFN (Italy); MEST, NRF, GSDC of KISTI, and WCU (Korea);
MNiSW (Poland); MES and RFAAE (Russia); ARRS (Slovenia);
SNSF (Switzerland); NSC and MOE (Taiwan); and DOE and NSF (USA).



\begin{thebibliography}{99}
\bibitem{bib:cc-disclaimer}
Throughout this paper, the inclusion of the charge-conjugate decay modes is implied unless otherwise stated.
\end{thebibliography}
\end{document}